\documentclass[a4paper,11pt]{article}
\usepackage{pos}
\usepackage{graphicx}
\usepackage[colorlinks=true
,urlcolor=posurl
,anchorcolor=posurl
,citecolor=posurl
,filecolor=posurl
,linkcolor=posurl
,menucolor=posurl
,linktocpage=true
,pdfa=true
]{hyperref}

\title{PrecisionSM: an annotated database for low-energy $e^+e^-$ hadronic cross sections}

\author*[a]{Lorenzo Cotrozzi}
\author[b,c]{Anna Driutti}
\author[a]{Fedor Ignatov}
\author[c,d]{Alberto Lusiani}
\author[a,c]{Graziano Venanzoni}

\affiliation[a]{University of Liverpool,\\
  Oxford St, Liverpool L69 7ZE, United Kingdom}

\affiliation[b]{University of Pisa, Department of Physics,\\
Largo B. Pontecorvo 3, Pisa, Italy}

\affiliation[c]{INFN - Sezione di Pisa,\\
Largo B. Pontecorvo 3, Pisa, Italy}

\affiliation[d]{Scuola Normale Superiore,\\
P.za dei Cavalieri, 7, Pisa, Italy}

\emailAdd{lorenzo.cotrozzi@liverpool.ac.uk}
\emailAdd{anna.driutti@unipi.it}
\emailAdd{F.Ignatov@liverpool.ac.uk }
\emailAdd{alberto.lusiani@sns.it}
\emailAdd{graziano.venanzoni@liverpool.ac.uk}

\abstract{PrecisionSM is an annotated database that compiles the available data on low-energy cross sections of electron-positron collisions into hadronic channels. This database organizes and collects data samples from $e^+e^-$ experiments, which are used as input for the data-driven theoretical evaluation of the muon anomalous magnetic moment, $a_{\mu}$, serving as a precise test of the Standard Model when compared to the experimental measurements of $a_{\mu}$. The database is accessible through a custom website~(\href{https://precision-sm.github.io}{https://precision-sm.github.io}) which contains details about the data samples, such as the treatment of radiative corrections, as well as links to papers on INSPIRE-HEP and to tables on HEPData.\\
The PrecisionSM database was developed within a Joint Research Initiative in the group application of the European hadron physics community, STRONG2020, and is now incorporated into the RadioMonteCarLow2 Working Group (RMCL2 WG) activities, which have the more general goal of improving the theoretical description of scattering processes at $e^+e^-$ colliders. The results of Phase I of the new RMCL2 WG have been published in Ref.~\cite{Aliberti:2024fpq}.\\
In this proceeding, we will report on the status of the PrecisionSM database, which currently contains a list of the dominant $2\pi$ channel as well as $3\pi$ and $\pi^0\gamma$, and on the ongoing work for the other channels and for responsive plots.}

\FullConference{The European Physical Society Conference on High Energy Physics (EPS-HEP2025)\\
7-11 July 2025\\
Marseille, France\\}

\begin{document}
\maketitle

\section{Introduction}

\noindent
The measurements of low-energy hadronic cross sections in $e^+e^-$ collisions are a key ingredient to probe the Standard Model (SM) of Particle Physics, and investigate any physical effects that are not yet explained by it. The $e^+e^-\to hadrons$ data can be used as input for the evaluation of the running of $\alpha_{em}$, the electromagnetic coupling constant, and also for the determination of the hadronic contribution to the anomalous magnetic moment of the muon, $a_{\mu}$, in the so-called \lq\lq data-driven dispersive approach\rq\rq. The leading order hadronic contribution to $a_{\mu}$ is obtained through the integral in Eq.~(\ref{eq:hvp_lo})~\cite{Aoyama:2020ynm, Aliberti:2025beg}:

\begin{align}\label{eq:hvp_lo}
    a_{\mu}^{\mathrm{HLO}}&=\left(\frac{\alpha m_{\mu}}{3\pi}\right)^2\int_{s_{\mathrm{thr}}}^{\infty}ds\frac{\hat{K}(s)}{s^2}R_{\mathrm{had}}(s)\\
    R_{\mathrm{had}}(s)&=\frac{3s}{4\pi\alpha^2}\sigma^0\left[e^+e^-\to\mathrm{hadrons(+\gamma)}\right],
\end{align}

\noindent
where $m_{\mu}$ is the muon mass, $\alpha\equiv e^2/(4\pi)$ is the fine structure constant, $\hat{K}(s)$ is a kinematic kernel function and $R_{\mathrm{had}}(s)$ is the hadronic $R$-ratio, i.e. the total $e^+e^-\to\mathrm{hadrons}$ cross section normalized to the Born cross section $e^+e^-\to\mu^+\mu^-$ in the point-like approximation. The superscript $0$ in $\sigma^0$ indicates the \emph{bare} cross section, \lq\lq undressed\rq\rq\ from Vacuum Polarization effects, to prevent double counting; in addition, the cross section used in the integral must be inclusive of final-state radiation of additional photons.

\noindent
In 2020, the Muon $g-2$ Theory Initiative recommended a theoretical value for $a_{\mu}$ that was based on the dispersive approach for the leading-order hadronic contribution, with a total precision of 370\,ppb~\cite{Aoyama:2020ynm}; this value was in $3.7\,\sigma$ tension with the experimental value for $a_{\mu}$ at the time~\cite{Muong-2:2006rrc}.
In 2021, the BMW collaboration published the first calculation of $a_{\mu}^{\mathrm{HLO}}$ with sub-percent precision~\cite{Borsanyi:2020mff} based on an alternative approach, namely \lq\lq lattice QCD\rq\rq. This result, when used for the calculation of $a_{\mu}$, was closer to the experimental value for $a_{\mu}$ and in $2.2\,\sigma$ tension with the value from the dispersive approach.
Moreover, in 2023, the CMD-3 collaboration released new results for the $e^+e^-\to\pi^+\pi^-$ cross section which disagreed with all the previous measurements used as input to the 2020 recommendation, and which were less in tension with the experimental value when used individually for the HLO dispersive approach~\cite{CMD-3:2023alj}.
In 2025, the Muon $g-2$ collaboration at Fermilab published the final experimental value for $a_{\mu}$, reaching the unprecedented precision of $124\,$ppb when combined with previous experimental results~\cite{Muong-2:2025xyk}; in the same year, the Muon $g-2$ Theory Initiative recommended a theoretical value for $a_{\mu}$~\cite{Aliberti:2025beg} which was based solely on lattice QCD for the evaluation of $a_{\mu}^{\mathrm{HLO}}$, and which implied no tension between the SM and experiment at the current level of precision. The persisting tensions within the $e^+e^-$ hadronic cross section measurements used for the dispersive approach, as well as the increased $a_{\mu}$ experimental precision, motivated renewed efforts to shed light on the discrepancies and further improve the accuracy of the dispersive approach.

\section{The STRONG2020 and RadioMonteCarLow2 activities}

\noindent
From 2006 to 2019, a community of experimentalists and theorists cooperated synergically within the RadioMonteCarLow Working Group (WG) with the goal of studying radiative corrections and Monte Carlo generators. The WG produced a final report~\cite{WorkingGrouponRadiativeCorrections:2010bjp} which contained an overview and future prospects of the experimental results and status of Monte Carlo generators for each of the five sections: high-precision luminosity measurements at low energies meson factories; aspects of the direct $R$ measurement performed at $e^+e^-\to hadrons$ experiments with energy scans; Initial State Radiation; $\tau$-lepton physics; calculation of vacuum polarization with emphasis on the hadronic contributions. The efforts of this WG were recently revived with the publication of new theoretical and experimental inputs to $a_{\mu}$. From 2020 to 2024, the European Union STRONG2020 Project (see~\href{http://www.strong-2020.eu/}{http://www.strong-2020.eu/}) incorporated parts of the RadioMonteCarLow WG, with the aim of studying strong interactions by combining knowledge from many frontiers: high and low energy physics, instrumentation and research infrastructures. The Project was concluded in 2024, and the activities of the former WG were incorporated in the newly established RadioMonteCarLow2 WG, which published its first report in 2024~\cite{Aliberti:2024fpq,Paltrinieri:EPS25proc}.

\section{PrecisionSM: the Precision Standard Model Database}

\begin{figure}[htbp]
\centering
\includegraphics[width=0.95\textwidth,alt={Screenshot of pipi table from the PrecisionSM database}]{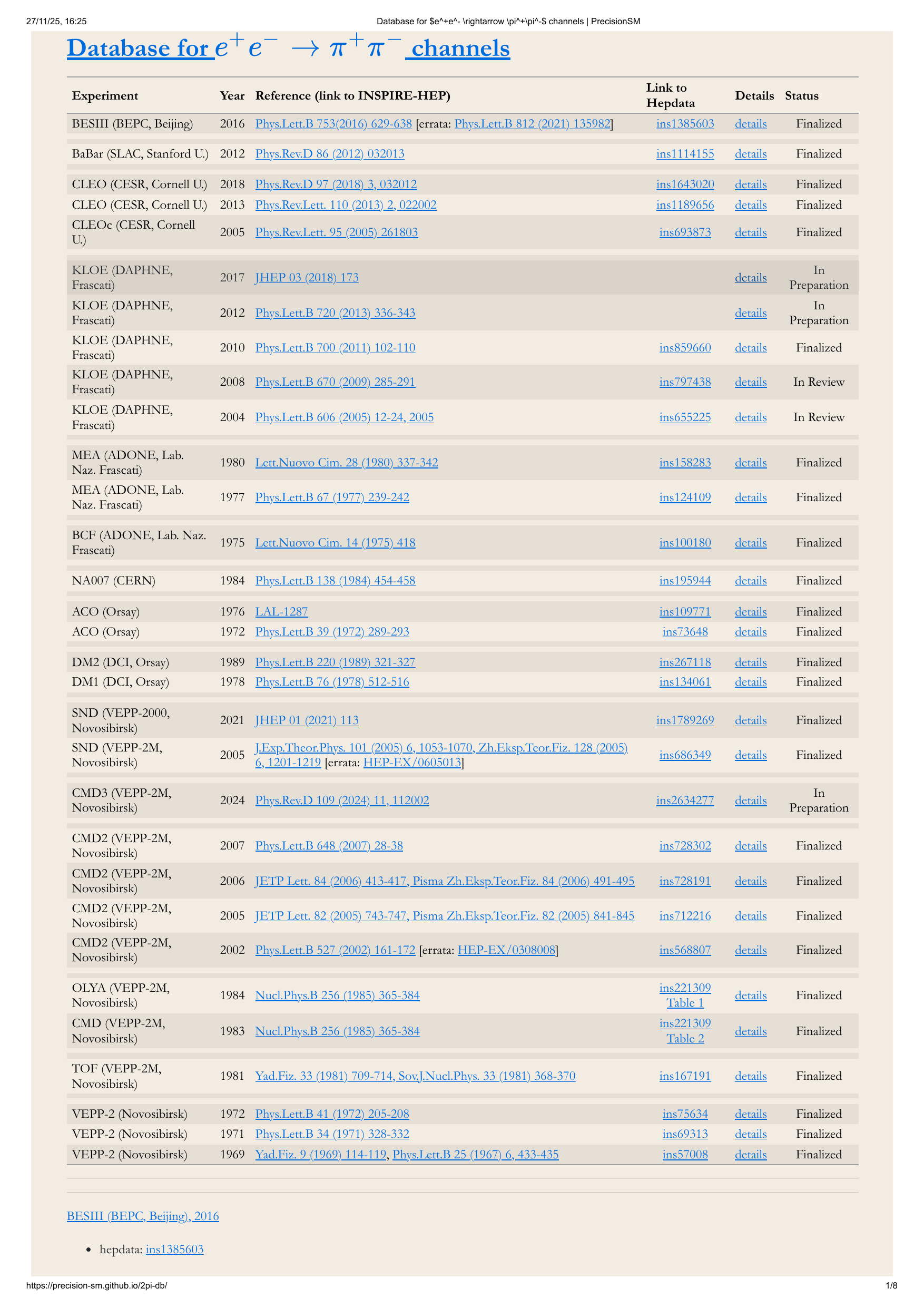}
\caption{Screenshot of the PrecisionSM table that lists all published measurements in the $e^+e^-\to\pi^+\pi^-$ channel. The up-to-date version of the table, accompanied by details on each experimental result, is accessible on the PrecisionSM website from the following link:~\href{https://precision-sm.github.io/2pi-db/}{https://precision-sm.github.io/2pi-db/}.\label{fig:table_2pi}}
\end{figure}

\noindent
Within the STRONG2020 Project, the PrecisionSM database started as a specific goal of the JRA3-PrecisionSM activity, which focused on the precise determination of SM parameters, such as: the anomalous magnetic moment of the muon, $a_{\mu}$; the CKM matrix element $V_{ud}$ from $\beta$ decay; the extraction of the weak mixing angle from parity-violating electron scattering~\cite{JRA3}.\\

\noindent
PrecisionSM is an annotated database for low-energy hadronic $e^+e^-$ cross sections available in literature, accessible through the following website:~\href{https://precision-sm.github.io}{https://precision-sm.github.io}, created by a Nikola static website generator~(\href{https://getnikola.com/}{https://getnikola.com/}), based on the files stored on the PrecisionSM GitHub repository. The database is divided by hadronic channels and for each channel it provides a list of the published experiments that measured the cross section, with annotated comments and links to the data.\\

\noindent
The PrecisionSM database is built in the following steps:

\begin{enumerate}
    \item \textbf{Data Collection.} In this phase, we make a list of all the experimental measurements that should be included in the input to the hadronic $R$-ratio, based on consultations with experts and on previously published reviews, such as the ones in Ref.~\cite{Aliberti:2025beg,Whalley:2003qr}. The list is divided into sublists based on the hadronic channel, and as a first step we started from the $\pi^+\pi^-$ final state, which accounts for most of the contribution to $a_{\mu}^{\mathrm{HLO}}$ through Eq.~(\ref{eq:hvp_lo}). The following collected channels were $\pi^+\pi^-\pi^0$ and $\pi^0\gamma$, recently posted on the website.
    \item \textbf{Upload of data in a public repository.} In this step, we communicate with the experimental collaborations to identify an expert point-of-contact. The expert submits the published data to the public repository HEPData~(\href{https://www.hepdata.net/}{https://www.hepdata.net/}) and appoints a reviewer, who will perform a cross-reference validation on the data based on the paper published on INSPIRE-HEP~(\href{https://inspirehep.net/}{https://inspirehep.net/}).
    \item \textbf{Catalogue data in an accessible way.} Once the entry on HEPData is validated and posted, it is then indexed on the PrecisionSM database. Fig.~\ref{fig:table_2pi} shows, as an example, the table for the $\pi^+\pi^-$ channel: each entry contains a link to the published paper on INSPIRE-HEP, to the data tables on HEPData, and the \lq\lq details\rq\rq\ link that points to notes on the bottom of the webpage. The additional information for each experiment includes: the range of energy for the cross section measurement; the treatment of radiative corrections and the Monte Carlo generators used; any other details on the experimental technique that might be of interest for any kind of comparisons among experiments.
\end{enumerate}

\begin{figure}[htbp]
\centering
\includegraphics[height=.35\textwidth,alt={First example of responsive plot on the PrecisionSM website}]{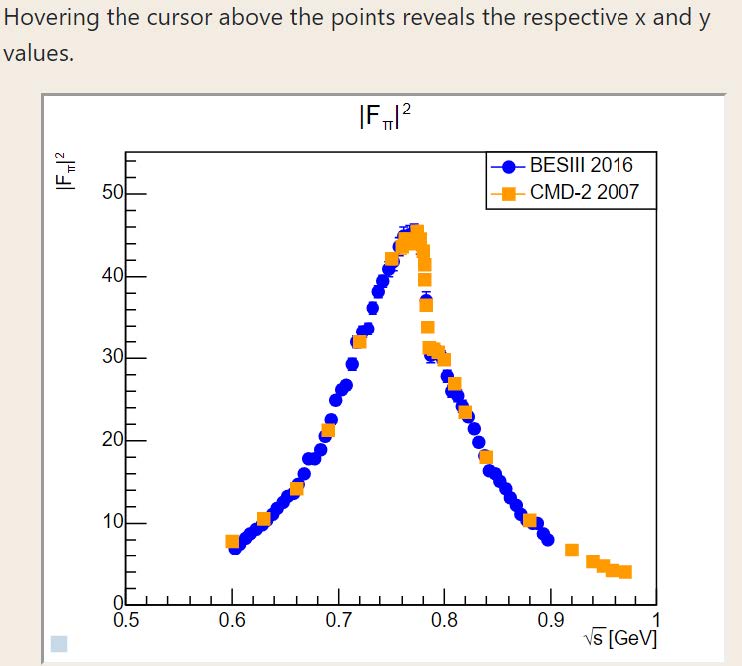}
\,
\includegraphics[height=.35\textwidth,alt={Second example of responsive plot on the PrecisionSM website; in preparation}]{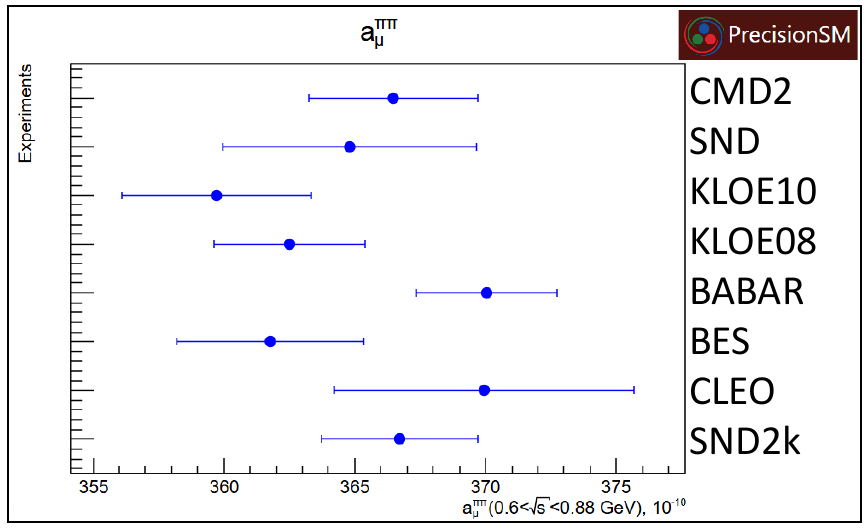}
\caption{Example of responsive plots of hadronic measurements stored in HEPData~\cite{PrecisionSM_plots}. The right-hand plot evaluates the integral of Eq.~(\ref{eq:hvp_lo}) for the $2\pi$ channel and is in preparation.\label{fig:responsive_plots}}
\end{figure}

\noindent
As a last step, the PrecisionSM website provides templates for tools to read from HEPData tables and prepare user-friendly responsive plots, at the link in Ref.~\cite{PrecisionSM_plots}, with examples shown in Fig.~\ref{fig:responsive_plots}. Currently, an additional plot is in preparation, that evaluates the $\pi^+\pi^-$ contribution to the integral of Eq.~(\ref{eq:hvp_lo}), for each of the published experiments that cover the energy range $[0.6,0.88]\,$GeV, with a calculation of the error bars that takes into account covariance matrices when provided.

\section{Conclusions}

\noindent
The STRONG2020 and RadioMonteCarLow(2) Working Groups have been facilitating the collaboration between the experimental and theoretical groups for 20 years, with the common goal of studying radiative corrections and Monte Carlo generators for low-energy hadronic $e^+e^-$ processes. Within these projects, the PrecisionSM activity is contributing with an annotated database for low-energy hadronic cross sections that compiles relevant information (e.g. Radiative Corrections treatment and systematic errors) for previously published $e^+e^-\to hadrons$ results, with the possibility to interact with responsive plots that take HEPData tables as an input. The database currently contains more than $60$ entries over three hadronic channels, and it will be expanded in order to list all the relevant channels. The information written in the PrecisionSM database are accessible to everyone and could be used, for example, by independent groups to calculate $a_{\mu}^{\mathrm{HLO}}$~\footnote{The authors of Ref.~\cite{Fowlie:2023cta} plan to use the database to continue their work on Treed Gaussian processes to model the $R$-ratio.}.
\noindent
All of these efforts have recently been revived by the measurement of the $e^+e^-\to\pi^+\pi^-$ cross section with the CMD-3 detector~\cite{CMD-3:2023alj} which is in tension with the previously published measurements, by the new sub-percent lattice QCD calculations of the leading-order hadronic contribution to the muon magnetic anomaly~\cite{Borsanyi:2020mff,Aliberti:2025beg}, and by the final measurement from the Fermilab Muon $g-2$ collaboration of the anomalous magnetic moment of the muon with unprecedented precision~\cite{Muong-2:2025xyk}.

\acknowledgments

\noindent
This work was supported by the European Union STRONG2020 project under Grant Agreement Number 824093, and by the Leverhulme Trust, \texttt{LIP-2021-014}.

\bibliographystyle{JHEP}
\bibliography{biblio}

\end{document}